\documentclass[fleqn,usenatbib]{mnras}
\usepackage{newtxtext,newtxmath}
\usepackage[T1]{fontenc}
\usepackage{ae,aecompl}
\usepackage{graphicx}	
\usepackage{amsmath}	
\usepackage{amssymb}	

\title[ASKAP Southern Sky Survey]{A Southern sky search for repeating Fast Radio Bursts using the Australian SKA Pathfinder.
}

\author[S. Bhandari et al.]
{S. Bhandari,$^{1}$\thanks{E-mail: shivani.bhandari@csiro.au},
K. W. Bannister$^{1}$,
C. W. James	$^{2}$,
R. M. Shannon$^{3, 4}$, \and
C. M. Flynn$^{4}$,
M. Caleb$^{5}$,
J. D. Bunton$^{1}$ \\
$^{1}$ CSIRO Astronomy and Space Science, PO Box 76, Epping, NSW 1710, Australia\\
$^{2}$International Centre for Radio Astronomy Research, Curtin University, Bentley, WA 6102, Australia \\
$^{3}$Centre for Astrophysics and Supercomputing, Swinburne University of Technology, Mail H30, PO Box 218, Hawthorn, VIC 3122, Australia\\
$^{4}$ARC Centre of Excellence for Gravitational Wave Discovery (OzGrav)\\
$^{5}$Jodrell Bank Centre for Astrophysics, School of Physics and Astronomy, The University of Manchester, Manchester M13 9PL, UK \\ }

\date{Accepted XXX. Received YYY; in original form ZZZ}
\pubyear{2018}

\begin{document}
\label{firstpage}
\pagerange{\pageref{firstpage}--\pageref{lastpage}}
\maketitle

\begin{abstract}
We have conducted a search for bright repeating Fast Radio Bursts in our nearby Universe with the Australian Square Kilometer Array Pathfinder (ASKAP) in single-dish mode. We used eight ASKAP 12-m dishes, each equipped with a Chequerboard Phased Array Feed forming 36 beams on the sky, to survey $\sim$30,000 deg$^{2}$ of the southern sky ($-90^{\circ} < \delta < +30^{\circ}$) in 158 antenna days. 
The fluence limit of the survey is 22 Jy~ms. We report the detection of FRB~180515 in our survey. We found no repeating FRBs in a total mean observation of 3~hrs per pointing divided into one hour intervals, which were separated in time ranging between a day to a month. Using our non-detection, we exclude the presence of a repeating FRB similar to FRB~121102 closer than $z=0.004$ in the survey area --- a volume of at least $9.4 \times 10^4$\,Mpc$^3$ --- at 95\% confidence.
\end{abstract}




\section{Introduction}
\label{sec:1}

\begin{figure}
    \includegraphics[scale=0.42]{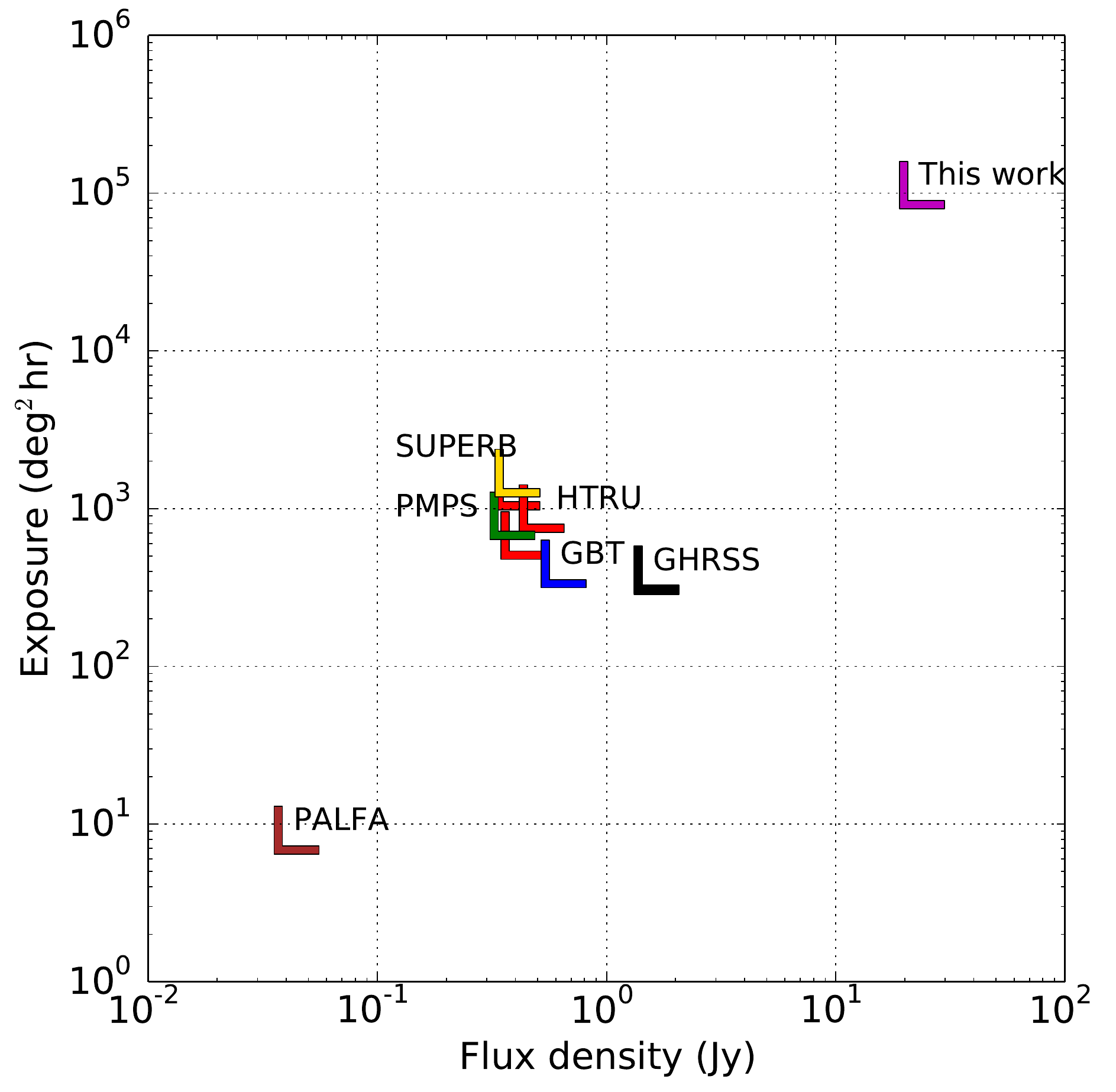}
    \caption{The phase space of sensitivity versus survey speed probed by current and archival transient surveys. The upper limits for various transient surveys are presented in this plot. The high, medium and low latitude High Time Resolution Universe Survey (HTRU) at Parkes are marked as red points \citep{BPSR}. Other surveys at the Parkes radio telescope including SUPERB \citep{SUPERB1} and PMPS \citep{Manchester2001} are shown as yellow and green points. Brown, blue and black data points presents the surveys conducted at the Arecibo \citep[PALFA]{Deneva2009,Patel2018}, the Green Bank telescope \citep[GNBCC]{Stovall2014} and the Giant Metrewave Radio telescope\citep[GHRSS]{Bhattacharyya2016}. The CRAFT all-sky survey using the ASKAP (this work) is presented using magenta data point, which explores a wide and shallow part of the phase space. }
\label{fig:phase space}
\end{figure}

The capability of instantaneously observing a large area of the sky has not only ramped up the detection of millisecond fast radio bursts (FRBs) but also diversified the current FRB population \citep{pbj+16}. FRBs have the capacity to be used as cosmological probes and offer an entirely new means of addressing questions such as the magnetic field of the intergalactic medium, the well-known problem of the ``missing baryons" \citep{Quinn,Ioka} and the dark energy density \citep{Zhou2014} by providing distances, when localised to distant galaxies and/or other sources, pending the nature of FRB hosts. The Australian Square Kilometer Array Pathfinder (ASKAP), a radio array consisting of 36$\times$12~m dishes and currently being commissioned in the Murchison Widefield Observatory have proved to be an excellent instrument for wide-field and blind searches for FRBs. It is equipped with Chequerboard Phased Array Feeds \citep[PAF]{Hay2008} making it a fast 21-cm survey machine. 

The Commensal Real-Time ASKAP Fast-Transients \citep[CRAFT]{mbb+10} latitude-50 survey has detected 20 new FRBs, surveying 
a large field of view in fly's eye mode ($150 - 270$~deg$^{2}$) \citep{Shannon2018}. The more luminous and relatively low dispersion measure sample of these ASKAP FRBs complements the distant events found in narrow-field and more sensitive surveys, e.g. \citet{Bhandari2018,SUPERB1}. 
These investigations of the brighter and closer FRB population are crucial in understanding the energetics of their radio emission. Additionally, detections with ASKAP permit robust measurements of the burst fluences because of the tight-packed beam configuration, whereas fluences obtained at other telescopes, such as Parkes, are generally lower limits.

Another advantage of ASKAP's high survey speed is the ability to re-observe the surveyed area, especially when it comes to fast transient searches including single pulse searches for pulsars. This can potentially lead to discoveries of new intermittent, nulling or sporadic sources. An extensive follow-up campaign of the FRB~121102, first detected in the PALFA survey by Arecibo in 2014 \citep{sch+14}, resulted in the discovery of the repeating pulses, making it the first repeating FRB \citep{ssh+16a}. Furthermore, its repeating nature led to an unambiguous localisation to a low-metallicity, dwarf host galaxy at $z = 0.19$ \citep{VLAlocalisation,Host}. Recently, six repeating pulses were detected from FRB~180814.J0422+73  \citep{R2}, one of the 13 FRBs reported by CHIME down to 400~MHz \citep{chime400}. The detection of the FRB~180814.J0422+73 hints at the existence of two classes of FRBs (repeating and non-repeating). Even though the spectral and temporal properties of the pulses observed from FRB~121102 and FRB~180814.J0422+73 are consistent with those observed for putative non-repeating FRBs \citep{Farah2018,RaviScience}, the extremely high rotation measure observed for FRB~121102 \citep{Michilli2018} differentiates the two speculated classes of FRBs.

The CRAFT collaboration has conducted an all-southern-sky survey with ASKAP leveraging its high survey speed and wide-field of view. In this paper we present the results our survey. In Section \ref{sec:2} we describe the observations, survey parameters and data reduction methods. We present our results including the discovery of FRB~180515 in Section \ref{sec:3}. Section \ref{sec:repeater} uses the non-detection of a repeating FRB to limit the number of objects similar to FRB~121102 in the local Universe. We describe our conclusions in Section \ref{sec:5}.

\begin{figure*}
	\includegraphics[scale=0.55]{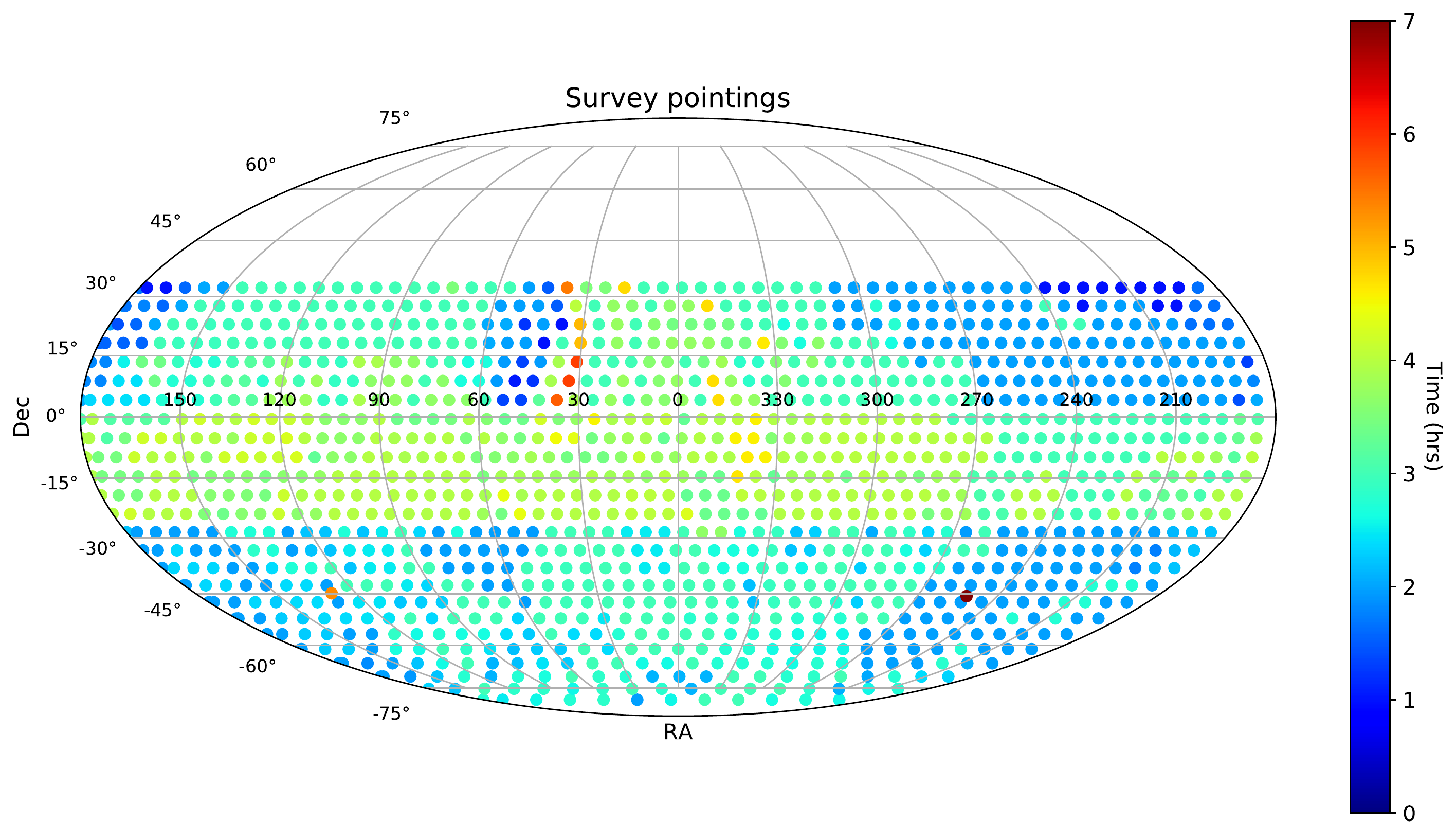}
    \caption{An aitoff projection of 1286 survey pointings covering $\sim 30000~$deg$^{2}$ of the southern sky. A pointing consists of a set of 36 beams arranged in a closepack pattern, observing a given position on the sky. 
    They are colour coded by their total observation time in hours. The typical total time spent on a given position is $\sim 3$ hours (see Figure  \ref{fig:times}) }
    \label{fig:all-sky pointings}
\end{figure*}

\section{Survey Overview.}
\label{sec:2}
The main goal of the CRAFT all-sky survey was to search for nearby bright repeating FRBs by re-observing the sky multiple times with different cadences. Figure \ref{fig:phase space} presents the phase space probed by high time resolution surveys conducted by different telescopes. The CRAFT all-sky survey with ASKAP explores a different region of this phase space, targeting the most energetic and/or closest FRBs over the broadest region of sky.

\subsection{Observations and Survey parameters}
\begin{figure}
	
    \includegraphics[scale=0.39]{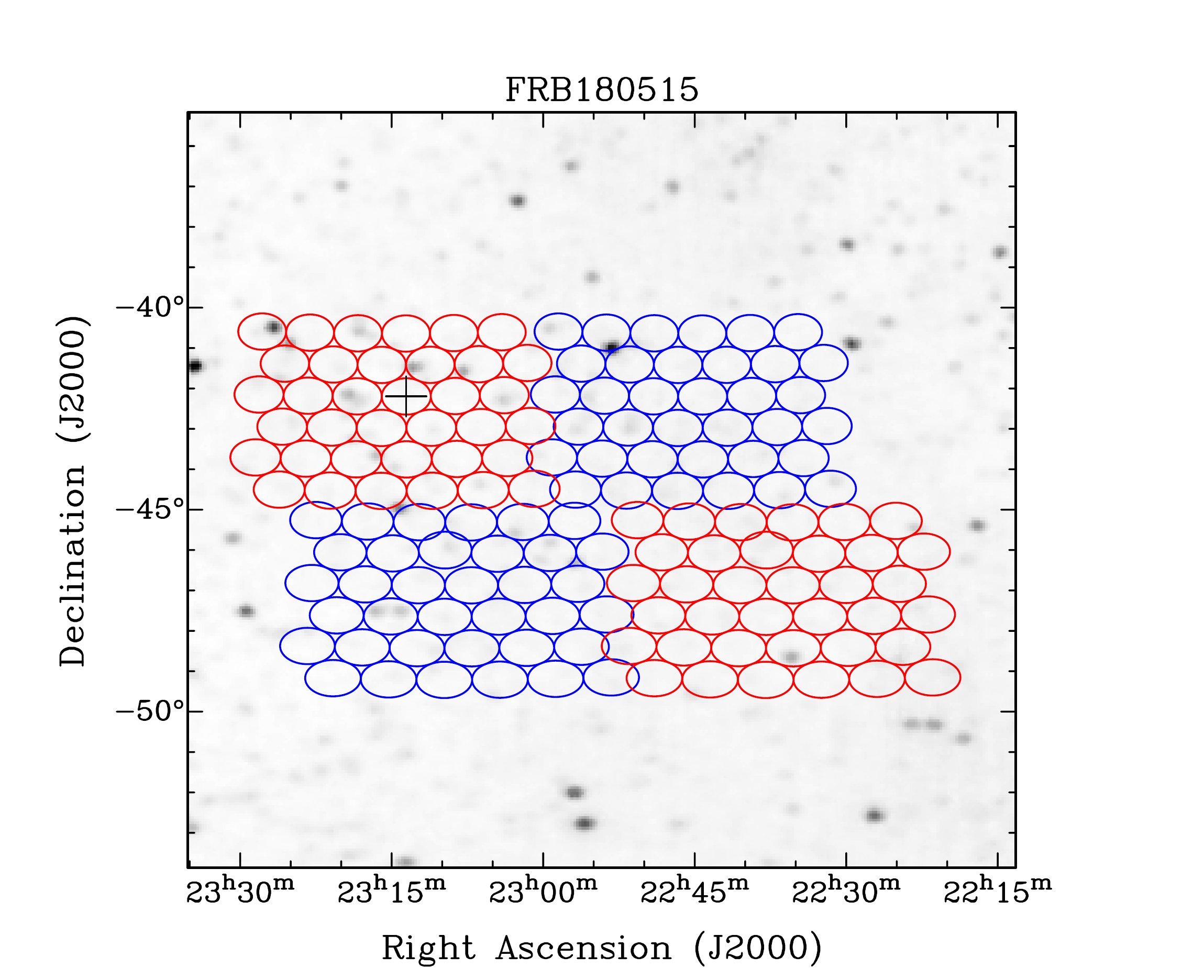}
    \caption{A plot of the closepack36 footprint arrangement of 36 beams on the sky for four pointings. Greyscale image is the CHIPASS 1.4~GHz radio continuum map, different colours represent different pointings, and the black cross represents the position of FRB~180515. The symbols represent the FWHM of the beam.}
    \label{fig:FRBbeam}
    \end{figure}

The observations were conducted in ASKAP's fly's eye mode, with each antenna pointing to a different position on the sky. The voltages measured by the PAFs are amplified, digitized and filtered into 336 coarse channels of 1~MHz width. The beam-former for each antenna constructs 36 beams by summing and weighting signals from individual PAF elements. Each beam has a FWHM of $\sim 1\rm~deg$ and 36 beams are overlapped to yield a field of view $\sim 30 \rm~deg^{2}$ at 1.4~GHz \citep{Hotan2014}. We have used eight ASKAP dishes, simultaneously observing $\sim 240 \rm~deg^{2}$ of the sky. The beams were arranged in a the ``closepack36" configuration with a pitch of $0.9^{\circ}$ and $60^{\circ}$ rotation on the sky. The survey covers the full sky visible to ASKAP in 1286 pointings, which we divided into three regions with declinations: $-80^{\circ}$ to $-30^{\circ}$ (AS1), $-30^{\circ}$ to $0^{\circ}$ (AS2) and $0^{\circ}$ to $30^{\circ}$ (AS3), also shown in Figure \ref{fig:all-sky pointings}.
The polarization axis (the roll axis of the dish) was set to $120^{\circ}$ and $-60^{\circ}$ for pointings north and south of declination $-30^{\circ}$ respectively, resulting in a beam pattern on the sky with minimum overlap, as shown Figure \ref{fig:FRBbeam}.

\begin{figure}
	\includegraphics[scale=0.42]{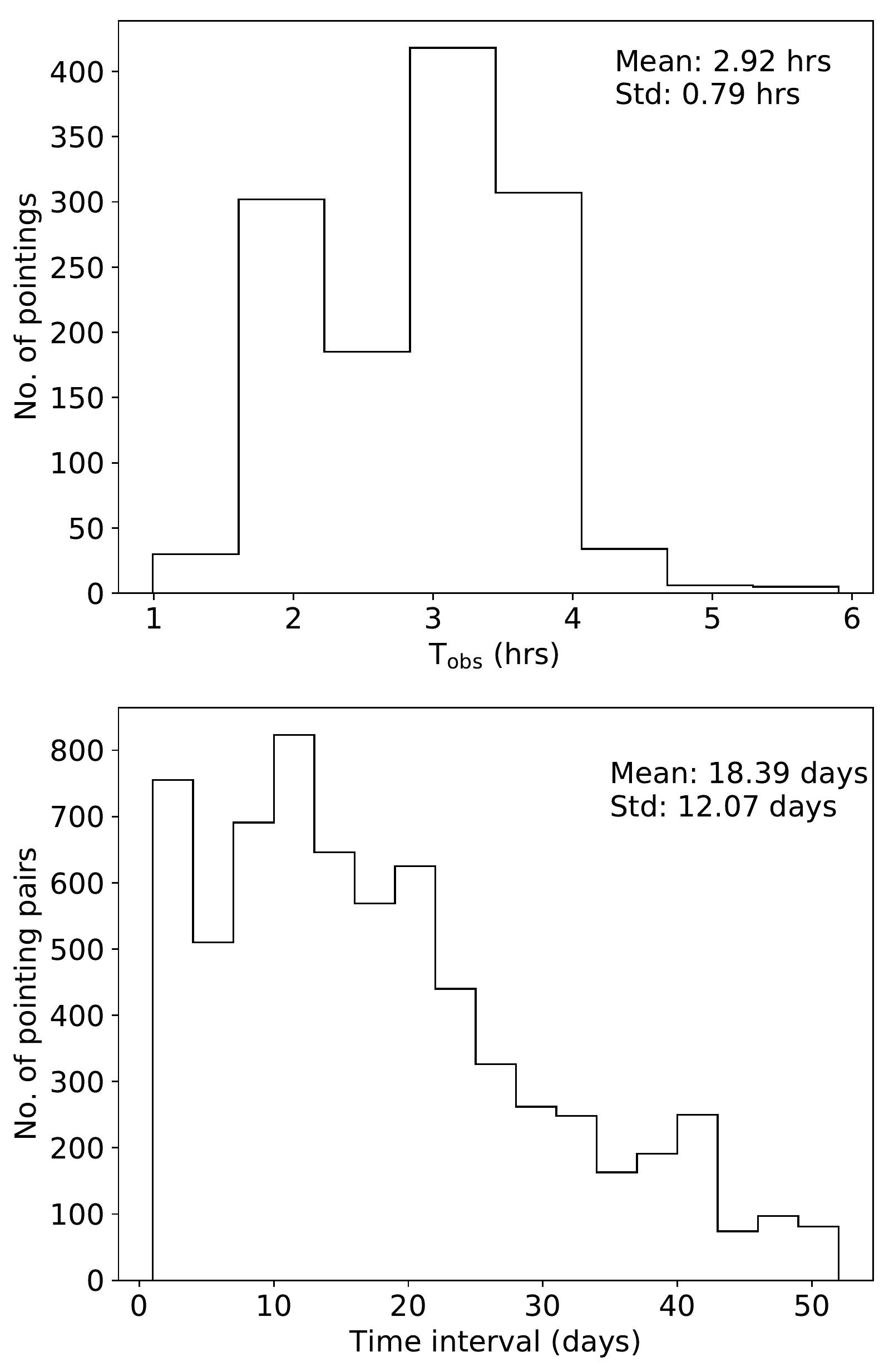}
    \caption{Top panel: Histogram of observation times for survey pointings. The mean observation time is 2.92~hrs. Bottom panel: Histogram of the time intervals between re-observation of a given pointing. This varies from a day to a month with a mean revisit time of 18 days.}
\label{fig:times}
\end{figure}

The observations were conducted between 2018 March 08 and 2018 May 21 with a total on-sky time of $\sim$158 antenna days. Each pointing was observed for a minimum duration of one hour. The mean total observation time per pointing for the survey was $\sim$3~hrs as presented in the top panel of the Figure \ref{fig:times}. The pointings were re-observed at least once; the revisit time varied from a day to a month, with a mean time of $\sim 18$ days as seen in the bottom panel of Figure \ref{fig:times}. The technical specifications of the survey are listed in Table~\ref{tab:survey_spects}.

The observation strategy was motivated by the clustered nature of the repeating pulses from FRB~121102 \citep{ssh+16b,2017ApJ...850...76L,2018ApJ...863....2G}. \citet{Oppermann2018} adopted a Weibull distribution to explain the observed clustering of pulses and also proved that observations interspersed with gaps have a higher chance of detecting a burst than a continuous observations with the same total time. 

Scaling from the results of \citet{Shannon2018}, who reported detection of 20 FRBs in $\sim$1300 antenna days, we expected to find $\sim$2.4 FRBs in our survey of 158 antenna days. One such FRB was found (FRB~180515), consistent with this estimate (see Section \ref{sec:discovery}).

\begin{table}
	\centering
	\caption{The specifications of the CRAFT All-Sky survey}
	\label{tab:survey_spects}
	\begin{tabular}{lcc} 
			\hline
        Region AS1  &$-80^{\circ} <  \delta   < -30^{\circ}$ \\
        Region AS2 &$-30^{\circ} < \delta < 0^{\circ}$ \\
        Region AS3 &$0^{\circ} < \delta  < 30^{\circ}$ \\
		$\tau_{\rm obs}$ \rm per pointing (sec) & 3600 \\
	SEFD (Jy)	&  $\sim$2000 \\
	Total pointings & 1286 \\
        $N_{\rm beams}$ per pointing & 36\\
		Bandwidth (MHz) & 336 \\
        $\tau_{\rm samp}$ (ms)& 1.265\\
		$\Delta \nu_{\rm chan}$ (MHz) & 1\\
        Centre Frequency (GHz)& 1.32\\
		$N_{\rm chans}$ & 336\\
 
		\hline
	\end{tabular}
\end{table}

\subsection{Data reduction pipeline}
The square-law-detected voltages from the beamformers are integrated to yield a time resolution of 1.265~ms. The filterbank data is then searched offline for bright single pulses using FREDDA (K. W. Bannister 2019, in preparation), the GPU-based implementation of the Fast Dispersion Measure Transform algorithm \citep[FDMT]{Zackay2017}. 
Each block of data is flagged and rescaled before performing the FDMT operation independently on each beam. 
The search for single pulses is performed in 4096 dispersion measure trials with DM ranging from $0-3763$~pc~cm$^{-3}$ and boxcar width trials in the range of $1-32$ samples. The known pulsar candidates are disregarded after comparing them with pulsar catalogue (Qiu et al, submitted to MNRAS) and candidates with S/N $>$12 and width $<12$~samples are classified as FRB candidates, which are manually inspected. A detailed description of the search pipeline is in \citet{ASKAP}.


\section{Results}
\label{sec:3}

We found one FRB in our survey, which is consistent with the expectation within the errors. No repeats of this FRB were found. 

\subsection{Detection of FRB~180515}
\label{sec:discovery}
\begin{figure}
	\includegraphics[scale=0.5]{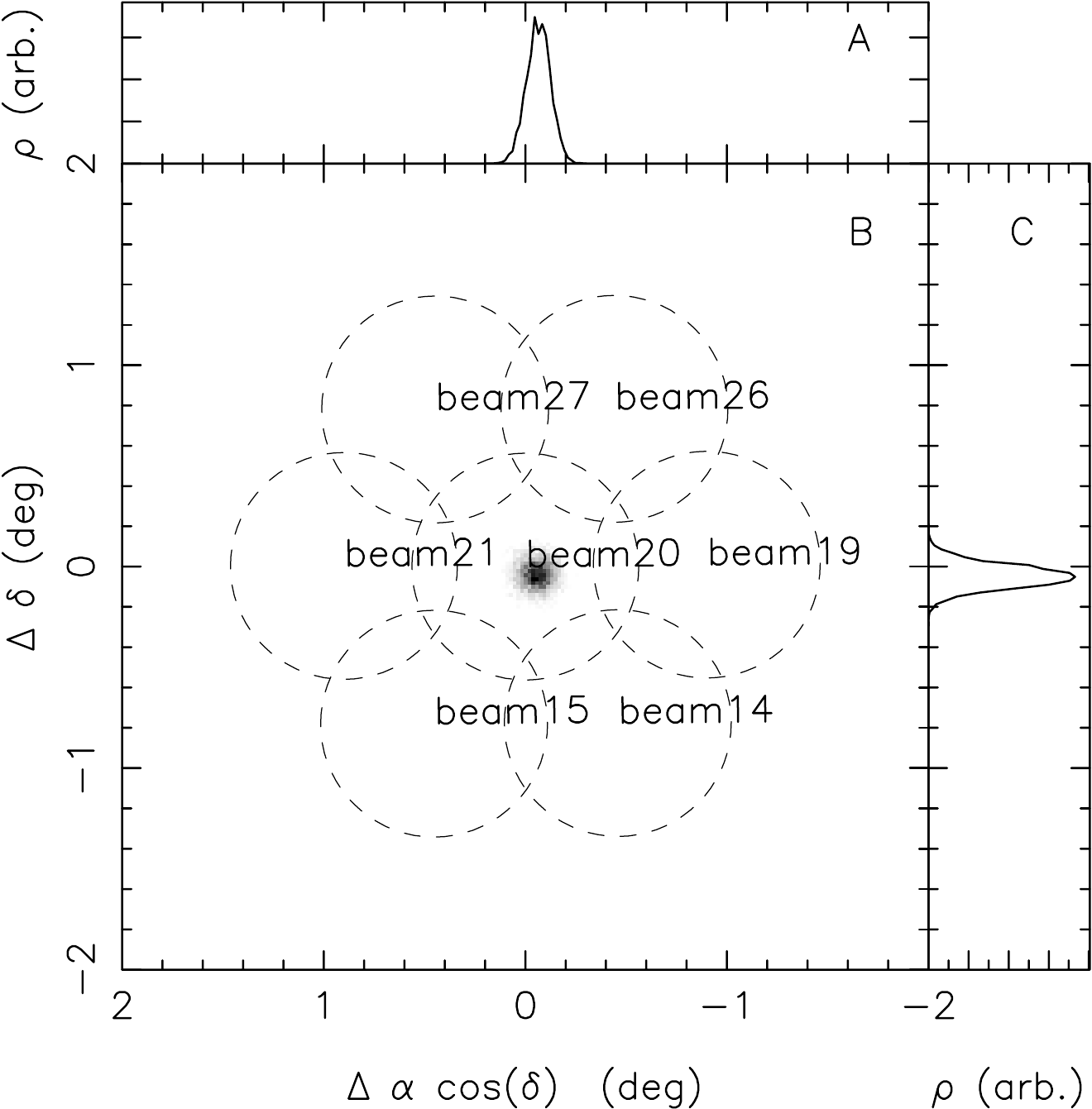}
	 \centering
    \caption{Constraints on position of FRB~180515. Panels A, B and C show the one and two-dimensional posterior distributions for burst position. The centre (0,0) in panel B is the best position of the FRB which is RA: 23:13:12 and DEC: $-$42:14:46, 4.9$'$ away from the beam centre. The marginal adjacent beam detections (S/N $< 3$) are stronger in beams 14, 15 and 19 than they are in beams 21, 26 and 27, hence the centroid of the posterior distribution has been shifted slightly in the direction of the former set of beams. This FRB has an positional uncertainty of a radius of 7$'$ at the 90$\%$ confidence level.}
     \label{fig:localisation}
\end{figure}

\begin{figure}
	\includegraphics[scale=0.6]{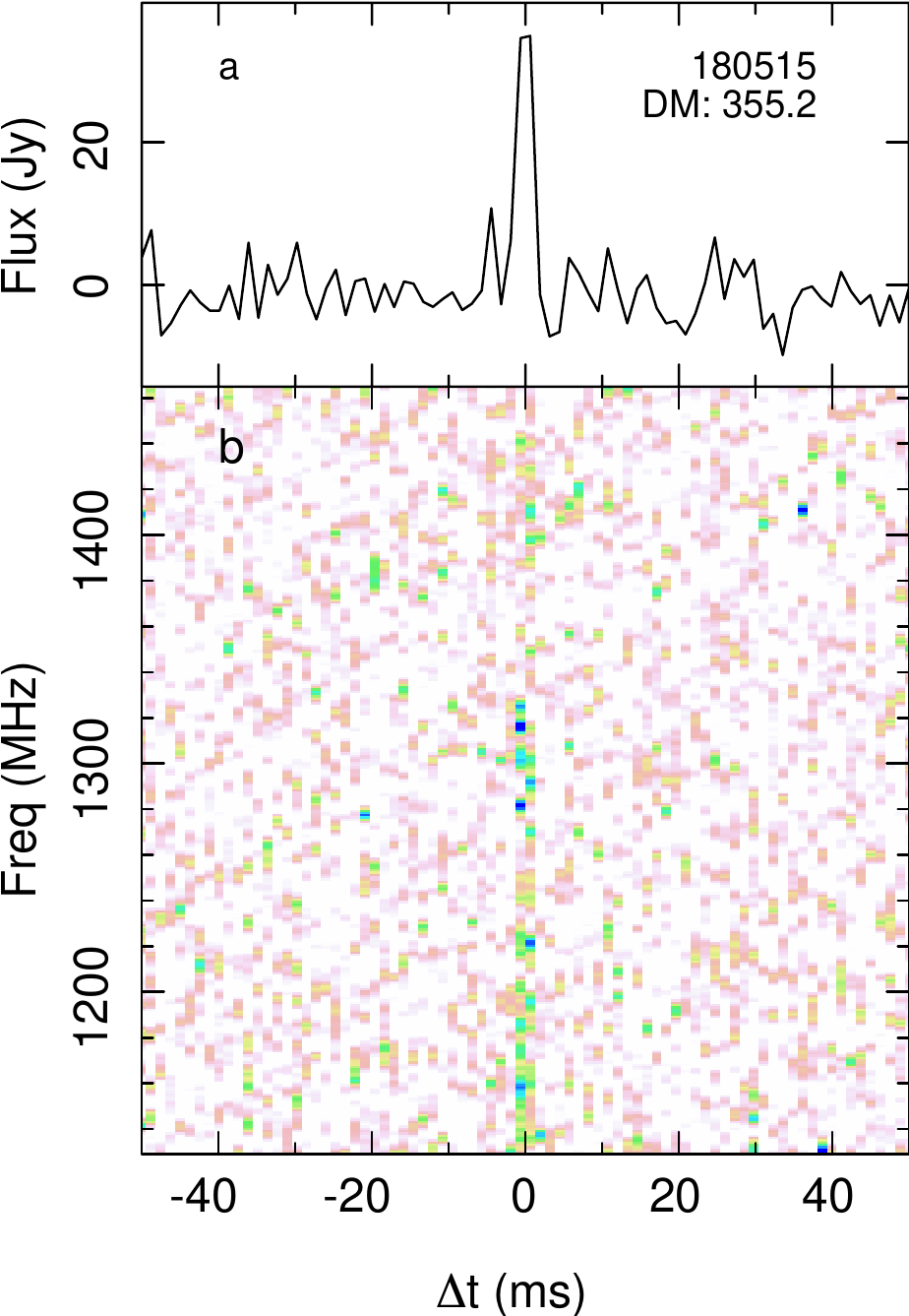}
    \centering
    \caption{Panel A: The pulse profile of FRB~180515 de-dispersed to a DM of 355.2~pc~cm$^{-3}$. We note a $3\sigma$ precursor spike, which is consistent with noise. Panel B: The FRB's dynamic spectrum. The colour map is set to range from the mean to 4$\sigma$ of the off-pulse intensity. The FRB is similar to other FRBs found at ASKAP at high Galactic latitude.}
   \label{fig:FRBprofile}
\end{figure}

FRB~180515 was discovered on 2018 May 15 UT 21:57:26.485 in the CRAFT AS1 region survey by antenna 15 after observing for 45 antenna days in the fly's eye mode. It was detected in beam 20 with a S/N of 12.1 at RA: 23:13:12 and DEC: $-$42:14:46, 4.9$'$ away from the beam centre, with the localisation error region of $7^{'}$ at the 90$\%$ confidence level. (Figure \ref {fig:localisation}). The burst has a dispersion measure (DM) of 355.2~pc~cm$^{-3}$ and is 1.9(4) ms wide. It has a fluence of 46(2) Jy~ms where the uncertainty in the position is accounted for in determining the uncertainty in the fluence. This is discussed in detail in \citet{ASKAP}. The burst is only marginally brighter in the lower half of the band  Figure \ref{fig:FRBprofile}. We have accounted for the offset position from the beam centre and the known gain variations in the band. This is discussed in the supplementary information of \citet{Shannon2018}. The burst does not show significant scattering and upper limit is presented in Table \ref{tab:frb_spects}.
 
\subsection{Maximum redshift and isotropic energy of FRB~180515.}
\begin{figure}
	\includegraphics[scale=0.35]{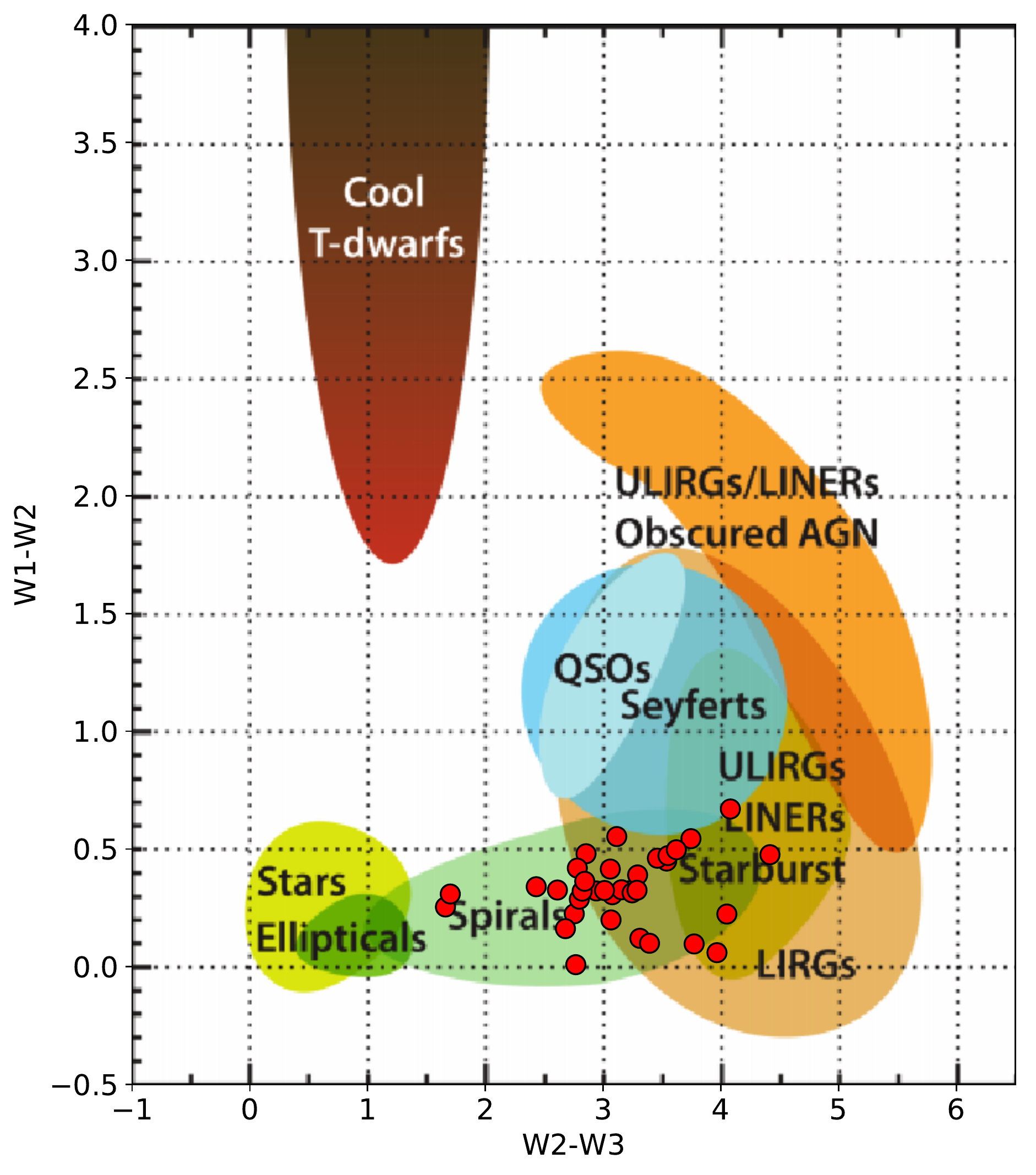}
    \caption{A plot of the WISE colours of 35 galaxies (red points) detected in WISExSCOSPZ catalogue in the FRB~180515 position error region, overlaid on the plot taken from \citet{Wright2010}. The galaxies reside in the spiral region of the plot. }
    \label{fig:WISE_galaxy}
\end{figure}

The total DM of an FRB (DM$_{\rm total}$) can be expressed as:
\begin{equation}
\rm DM_{total} = DM_{MW(disk)} + DM_{MW(halo)} + DM_{IGM} + DM_{Host}
\end{equation}
where DM$_{\rm MW(disk)}$ is the DM contribution from the Milky Way disk, DM$_{\rm MW(halo)}$ is the Milky Way halo contribution, DM$_{\rm IGM}$ is contribution from the intergalactic medium and DM$_{\rm Host}$ is the FRB host galaxy contribution. Using the Galactic models of NE2001 \citep{Cordes} and YWM16 \citep{YMW16}, we obtain a Milky Way disk DM contribution of $\sim$33~pc cm$^{-3}$ and $\sim 19$~pc cm$^{-3}$ respectively along the line of sight for FRB~180515. We assumed Milky Way halo and host galaxy DM contribution to be $\sim$12~pc cm$^{-3}$ and $\sim 45$~pc cm$^{-3}$ \citep{Mahony2018,Xu2015}. This results in an intergalactic medium contribution in the range $265<  \rm DM_{IGM} < 280$ pc~cm$^{-3}$. For a given excess DM, assuming the likelihood of $ \rm DM_{IGM}$ along a sightline to be a Gaussian with redshift-evolving mean and variance and a homogeneous prior redshift distribution for FRBs, the most probable redshift can be calculated using the methods described in \citet{Walker2018}. For $265<  \rm DM_{IGM} < 280$ pc~cm$^{-3}$, this redshift range is $0.255 < z_{\rm max} < 0.269$. A search in the WISExSCOS Photometric Redshift Catalogue \citep[WISExSCOSPZ]{Bilicki2016} in the localization region with redshift constrained to $z < 0.27$ resulted in 35 galaxies brighter than the survey magnitude limit $I = 18.5$. The galaxies have redshifts in the range of $0.04 - 0.259$ and they reside in the region of non-AGN ``spirals" in the WISE colour-colour plot \citep{Wright2010} as shown in the Figure \ref{fig:WISE_galaxy}.
However, we lack confidence in associating any galaxy to FRB~180515.

The in-band isotropic energy of an FRB using $\Lambda$CDM cosmology \citep{2016A&A...594A..13P} can be expressed as:

\begin{equation}
	E(J)= \frac{\mathcal{F}_{\rm obs} \times \rm BW \times 4 \pi D_{\rm L}^{2}
	 \times 10^{-29}}{(1+z)^{2-\alpha}} \label{eq:3}
	\end{equation}
    where $\mathcal{F}_{\rm obs}$ is the observed fluence for an FRB in Jy\,ms, BW is the bandwidth at ASKAP in Hz, $D_{\rm L}$ is the luminosity distance in metres, $z$ is the 
    redshift, and $\alpha$ is the spectral index of the source ($F \propto \nu^{\alpha}$). We assume a flat spectrum for FRBs i.e., $\alpha = 0$. 
	
Using equation \ref{eq:3} for FRB~180515, we obtain an isotropic energy of $1.6 \times 10^{33}$~J. These properties are also listed in Table~\ref{tab:frb_spects}.

\begin{table}
	\centering
	\caption{The measured and model dependent properties of FRB~180515.}
	\label{tab:frb_spects}
    \begin{tabular}{lcc}
      \hline
      \textbf{Measured Properties}  \\
      \hline
  
      Event time at 1.4~GHz UTC & 2018-15-05 21:57:26.485  \\
      ASKAP beam  & 20   \\
      Beam centre (Ra, Dec) (J2000) & 23:13:33.8, $-$42:11:51.3 \\
      FRB (Ra, Dec) (J2000) &23:13:12, $-$42:14:46   \\
      Localisation error & 7$'$ radius \\
       Galactic coordinates ($\ell$, $b$) & $349.5^{\circ}$, $-64.9^{\circ}$  \\
      Signal to noise ratio, (S/N) & 12.1  \\
      Dispersion measure, DM (pc cm$^{-3}$) & $355.2(5)$  \\
      Fitted width (ms) & $1.9(4)$   \\
      Scattering time (ms) & $ < 0.38^{+0.10}_{-0.12}$   \\
      Measured fluence (Jy ms) &  $46(2)$ \\
      \hline
    \textbf{Model-dependent properties}  \\
      \hline
      DM$_{\rm NE2001}$ (pc cm$^{-3}$)& $\sim33$  \\
      DM$_{\rm YWM16}$ (pc cm$^{-3}$)& $\sim19$  \\
      Max. inferred $z$             &      0.2 \\
      Max. comoving distance (Gpc)       &          0.9 \\
      Max. luminosity distance (Gpc)     &       1.1 \\
      Max. isotropic energy (10$^{33}$ J) &       1.6\\
      
      \hline
      \end{tabular}
\end{table}


\section{Limits on repeating FRBs in the local Universe}
\label{sec:repeater}

No repeat FRBs were found in the survey. The observed rate of FRBs from a population of progenitors producing repeat bursts can be described in terms of intrinsic progenitor properties --- in particular, the repetition rate $R$ as a function of burst energy --- and the population distribution, which we characterise as the redshift-dependent number density per co-moving cubic Mpc, $\rho(z)$. Given the exceptionally broad range of parameter space that could describe a repeating FRB (rate, time distribution, pulse strength, frequency structure, etc.), we use the first known repeater, FRB~121102, to characterise all potentially repeating FRBs, and use our observations to limit the density of similar objects in the local Universe. Knowledge of the second known repeater, FRB~180814.J0422+73 \citep{chime400,R2}, remains too rudimentary to use this object as a benchmark.
The statistical analysis of this object is described in \citet{James2019} which primarily uses the results of \citet{2017ApJ...850...76L} and \citet{2018ApJ...863....2G} to model the cumulative brightness distribution of bursts via: 

\begin{eqnarray}
R(E>E_0) & = &  R_0 \left( \frac{E}{E_0} \right)^{\gamma} \label{eq:intrinsic_rate} \\
R_0 & = & 7.4\,{\rm day}^{-1} \nonumber \\
E_0 & = & 1.7 \cdot 10^{38}\,{\rm erg}  \nonumber \\
\gamma & = & -0.9
\end{eqnarray}
for burst energy $E$, time-averaged rate $R$, and index of the power law slope of burst energies, $\gamma$. This is broadly consistent with the model of intrinsic properties presented by \citet{2017ApJ...850...76L}.

To examine the dependence of resulting limits on FRB properties, the limits are recalculated with $R_0$ reduced ten-fold, i.e.\ $R_0=0.74$\,day$^{-1}$.

The arrival time distribution of bursts from FRB 121102 is clearly non-Poissonian in time and has been modeled as a Weibull distribution by \citet{Oppermann2018} based on 80~hr of observations during which only 17 pulses were detected. However, more recently \citet{2018ApJ...866..149Z} show that the arrival time statistics maybe indeed be more Poissonian than previously though to be, especially during the `active' state of the source. We note that observational biases may play a role in skewing our perception of its activity.\footnote{For instance, the only article reporting a non-detection at radio wavelengths appears as a Research Note \citep{2018RNAAS...2a..30P}.} Even with a Weibull distribution that shows correlated arrival times with Arecibo, multiple regular short observations reduce a clustered repeating source to a near-Poisson process.

Here, we consider only the case of Poissonian arrival statistics, and refer readers to \citet{2016MNRAS.458L..89C} for a more extensive discussion of how non-Poissonian statistics will interact with the time-distribution of our observations reported in Figure \ref{fig:times}. 
Naively characterising FRB 121102 as being active or inactive (e.g.\ the `early 2016' vs `late 2016' periods of \citet{2017ApJ...850...76L}, or the final hour versus the first five hours of \citet{2018ApJ...863..150S}), with some probability $p_{\rm active}$ over the timescales reported in Figure \ref{fig:times}, will weaken our resulting limits by a factor of $1/p_{\rm active}$.

\subsection{Deriving limits}
\label{sec:limits}

We place limits on the population of repeating FRBs using our non-observation of repeated pulses, using the method of \citet{James2019}. The number of observed single FRBs --- here, one --- can also be used to limit the number of repeating sources, e.g.\ in the case of many objects repeating at low rates. Using single pulses will always result in a more powerful method, simply because it uses more information.

The former method is more robust however, in that it is not affected by the presence of a population of cataclysmic sources in addition to repeating sources, at least until the observed burst rate is so high that there is a non-zero probability of a chance coincidence of FRBs from the same location with the same DM.
\begin{figure*}
\begin{tabular}{cc}
\includegraphics[width=8cm]{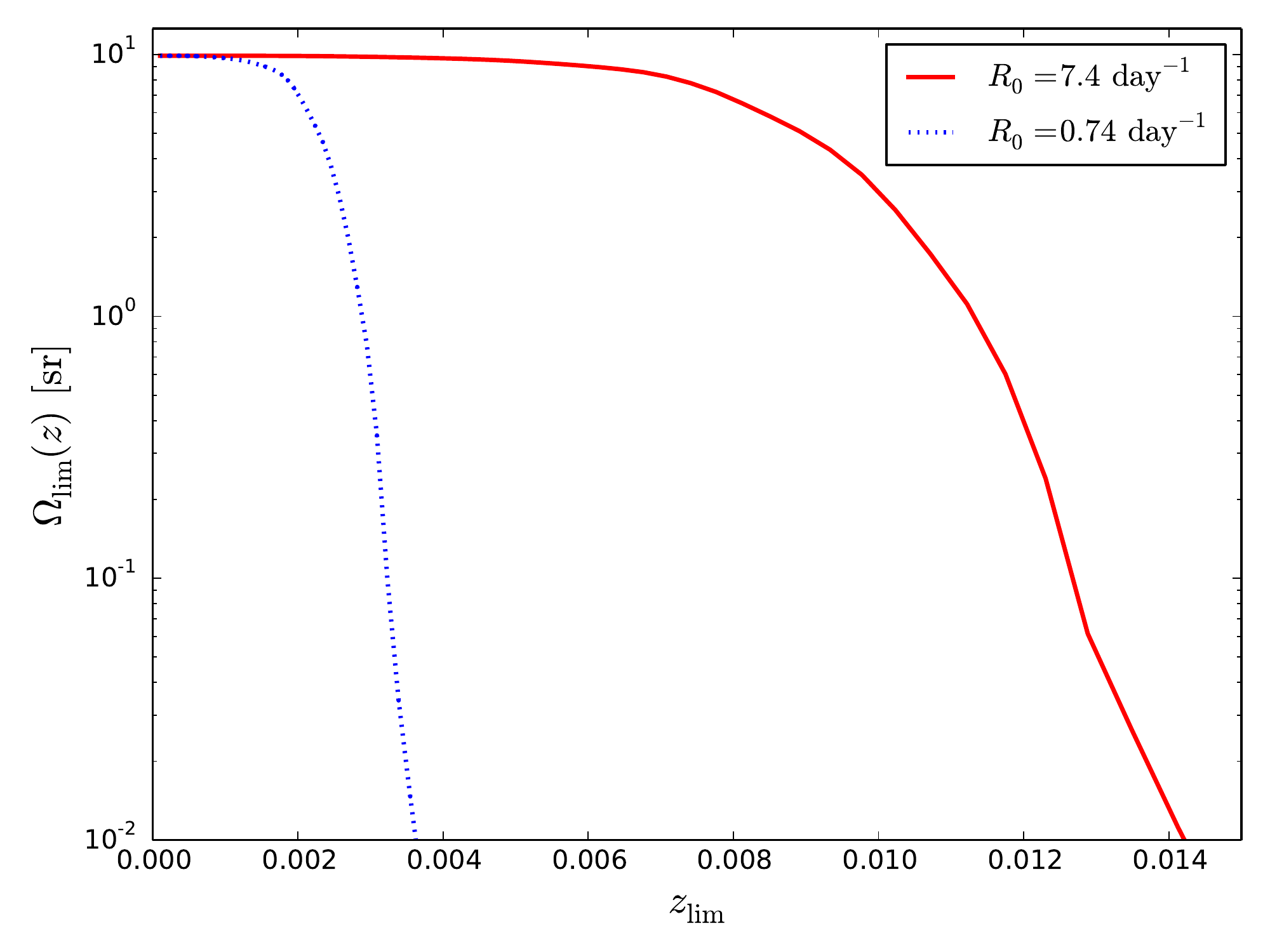} &
\includegraphics[width=8cm]{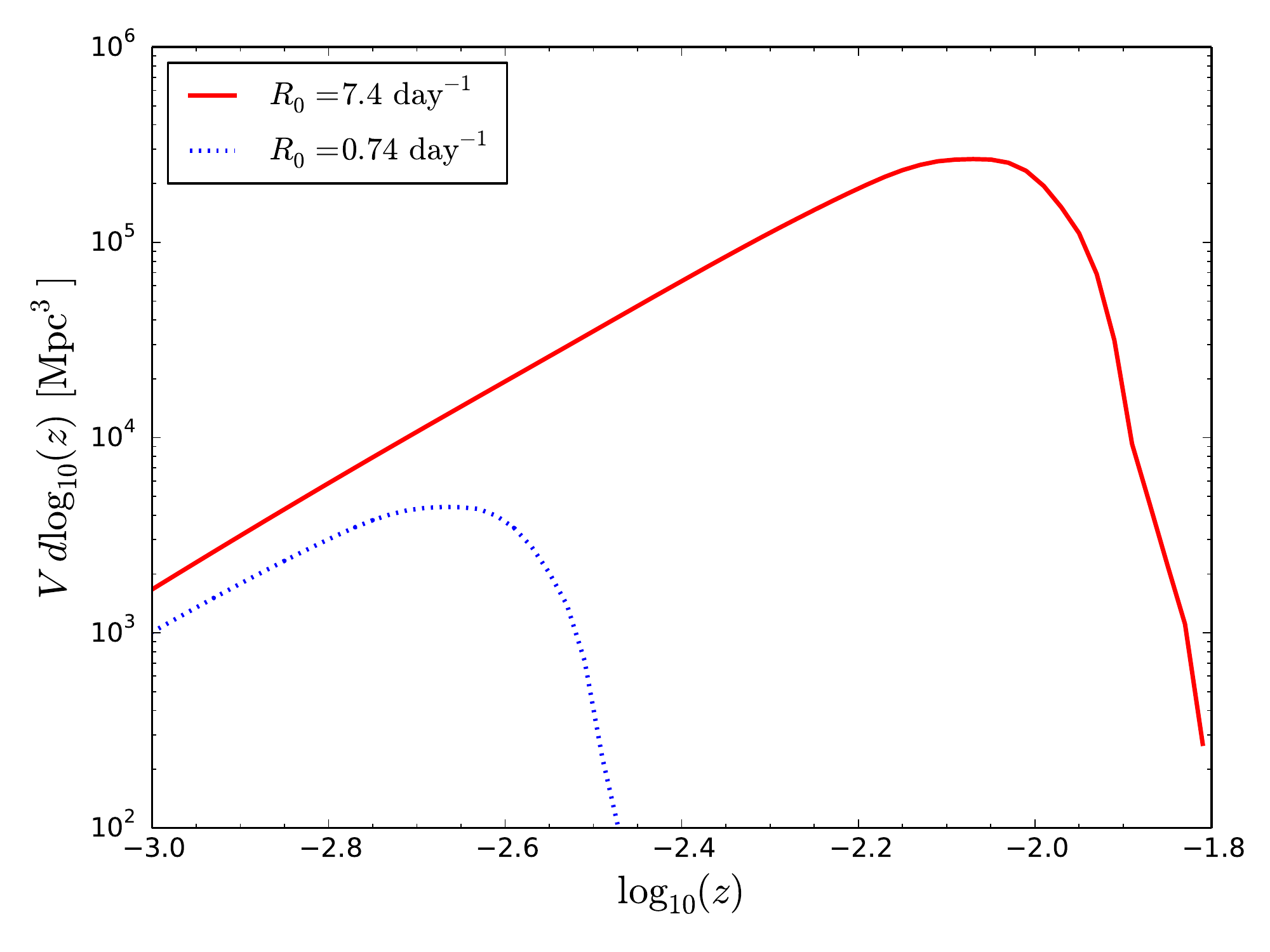} \\
\end{tabular}
\caption{Left panel: Solid angle $\Omega_{\rm lim}(z)$ over which the presence of a repeating FRB similar to FRB~121102 ($R_0=7.4$\,day$^{-1}$), and one with reduced rate $R_0=0.74$\,day$^{-1}$,  within redshift $z_{\rm lim}$ can be excluded at 95\% confidence level  (C.L.). Right panel: The differential volume in which the presence of a repeating FRB can be excluded at 95\% confidence level (C.L.) as a function of redshift, $z$.
}
\label{fig:plots_zlim}
\end{figure*}

Following \citet{James2019}, if a time $T_{\rm obs}$ is spent on-source, then the required intrinsic rate $R_{\rm lim}$ to exclude the presence of a repeater is given by:
\begin{eqnarray}
R_{\rm lim} & = & (1+z) \frac{\lambda_{\rm lim}}{T_{\rm obs}}
\end{eqnarray}
where the factor of $(1+z)$ is due to the time-dilation effects of redshift and $\lambda_{\rm lim} = 4.84$ is a Poissonian expectation value for an observation of 2 or more bursts from a repeating object at $95$\% confidence.

The value of $E_{\rm lim}$, above which the repetition rate is $R_{\rm lim}$, can be found by inverting equation \ref{eq:intrinsic_rate}:
\begin{eqnarray}
E_{\rm lim} & = & E_0 \left(\frac{R_{\rm lim}}{R_0}\right)^{\frac{1}{\gamma}} \nonumber \\
& = & E_0 (1+z)^{\frac{1}{\gamma}} \left(\frac{\lambda_{\rm lim}}{T_{\rm obs} R_0}\right)^{\frac{1}{\gamma}}.
\end{eqnarray}
However, intrinsic pulses of strength $E_{\rm lim}$ must be observable, being a function of $F$ and $z$. Given a threshold $F_{\rm th}$, there will exist a unique $z_{\rm lim}$ satisfying the condition that pulses of intrinsic strength $E_{\rm lim}$ are just observable at threshold $F_{\rm th}$ when generated at a redshift of $z=z_{\rm lim}$.

Using equation \ref{eq:3} to substitute $E_{\rm lim}$ for an expression involving $z$ and $F_{\rm th}$, we find:
\begin{eqnarray}
(1+z_{\rm lim})^{\alpha-2-\frac{1}{\gamma}} D_L^2(z_{\rm lim}) & = & \frac{E_0}{4 \pi F_{\rm th} \rm BW_{\rm obs}} \left(\frac{\lambda_{\rm lim}}{T_{\rm obs} R_0}\right)^{\frac{1}{\gamma}}. \label{eq:zlim}
\end{eqnarray}
where BW$_{\rm obs}$ is the observation bandwidth (here, 336\,MHz).

Using the closepack36 beam sensitivities from \citet{2018arXiv181004356J} to calculate $F_{\rm th}$ as a function of solid angle $\Omega$, equation \ref{eq:zlim} can be used to reduce the variables of $T_{\rm obs}$ and direction-dependent beam sensitivity to a single function, $\Omega(z_{\rm lim})$, describing the solid angle $\Omega$ in which the survey would have detected a repeating FRB at a given level of confidence for all $z \le z_{\rm lim}$. The total solid angle $\Omega_{\rm lim}(z)$ and volume $V_{\rm lim}$, over which the presence of a repeating FRB can be excluded are given by \citet{James2019}:
\begin{eqnarray}
\Omega_{\rm lim}(z) & = & \int_{z_{\rm lim}=z}^{\infty} \Omega(z_{\rm lim}) d z_{\rm lim}
\label{eq:solidlim}
\end{eqnarray}
and
\begin{eqnarray}
V_{\rm lim} & = & \int_0^{\infty} \Omega_{\rm lim}(z) D_H \frac{(1+z)^2 D_A^2}{E(z)} dz. \label{eq:vlim}
\end{eqnarray}
The integrand in equation \ref{eq:vlim} represents the comoving volume element, where $D_A$ is the angular diameter distance, $D_H$ the Hubble distance, and $E(z)$ represents the Hubble parameter as a function of redshift. 
The results are shown in Figure \ref{fig:plots_zlim}. The left panel  
shows the solid angle of sky in steradians as a function of the limiting redshift $z_{\rm lim}$,
and allows us to exclude the presence of a repeating FRB with similar properties to FRB~121102 and south of declination $\delta = +30^{\circ}$ within $z<0.004$ at 95\% confidence.
The right panel shows the results of integrating equation \ref{eq:vlim}, to yield the 95\% confidence limit on the surveyed volume as a function of redshift. We exclude the presence of a repeating FRB with properties similar to FRB~121102 in a volume, $V_{\rm lim}$ of not less than $9.4 \times 10^4$\,Mpc$^3$.

For FRBs repeating at different rates, the limited volume varies approximately as $(R_0^\frac{-1.5}{\gamma})$ (in Euclidean space). For $R_0=0.74$\,day$^{-1}$, it is $1.6 \times 10^3$\,Mpc$^3$.

\section{Conclusions}
\label{sec:5}
We have conducted an all-sky survey with ASKAP to search for bright repeating FRBs in our local Universe. We observed each field for atleast one hour and re-observed a given field mostly three times with cadence ranging from a day to a month. We found an FRB~180515 in our AS1 survey which has similar spectral features as other sample of ASKAP FRBs. However, we did not find any repeating FRBs. We used this non-detection to exclude the presence of FRB~121102 like sources within $z=0.004$ at 95\% confidence over the entire survey region, and calculated a limiting volume of $9.4 \times 10^4$\,Mpc$^3$, assuming the arrival times of FRB~121102 follow Poisson statistics. The limits are relatively weaker for more general distributions, such as Weibull.

\section*{Acknowledgements}

The Australian SKA Pathfinder is part of the Australia
Telescope National Facility which is managed by CSIRO.
Operation of ASKAP is funded by the Australian Government
with support from the National Collaborative Research
Infrastructure Strategy. ASKAP uses the resources of the
Pawsey Supercomputing Centre. Establishment of ASKAP, the
Murchison Radio-astronomy Observatory and the Pawsey
Supercomputing Centre are initiatives of the Australian
Government, with support from the Government of Western
Australia and the Science and Industry Endowment Fund. We
acknowledge the Wajarri Yamatji people as the traditional
owners of the Observatory site. SB would like to thank the anonymous referee for their comments and suggestions. SB would also like to thank Wael Farah for the scattering fit of the reported FRB and Charlie Walker for estimating the most probable redshift for a given DM contribution from the IGM. KB and RMS acknowledge the support of the Australian Research Council through grant DP18010085. RMS acknowledges salary support from the ARC through grants FL150100148 and CE17010004. CWJ acknowledges the support of the Australian Research Council Centres of Excellence for All Sky Astrophysics (CAASTRO, CE110001020)




\bibliographystyle{mnras}

\begin{thebibliography}{}
\makeatletter
\relax
\def\mn@urlcharsother{\let\do\@makeother \do\$\do\&\do\#\do\^\do\_\do\%\do\~}
\def\mn@doi{\begingroup\mn@urlcharsother \@ifnextchar [ {\mn@doi@}
  {\mn@doi@[]}}
\def\mn@doi@[#1]#2{\def\@tempa{#1}\ifx\@tempa\@empty \href
  {http://dx.doi.org/#2} {doi:#2}\else \href {http://dx.doi.org/#2} {#1}\fi
  \endgroup}
\def\mn@eprint#1#2{\mn@eprint@#1:#2::\@nil}
\def\mn@eprint@arXiv#1{\href {http://arxiv.org/abs/#1} {{\tt arXiv:#1}}}
\def\mn@eprint@dblp#1{\href {http://dblp.uni-trier.de/rec/bibtex/#1.xml}
  {dblp:#1}}
\def\mn@eprint@#1:#2:#3:#4\@nil{\def\@tempa {#1}\def\@tempb {#2}\def\@tempc
  {#3}\ifx \@tempc \@empty \let \@tempc \@tempb \let \@tempb \@tempa \fi \ifx
  \@tempb \@empty \def\@tempb {arXiv}\fi \@ifundefined
  {mn@eprint@\@tempb}{\@tempb:\@tempc}{\expandafter \expandafter \csname
  mn@eprint@\@tempb\endcsname \expandafter{\@tempc}}}

\bibitem[\protect\citeauthoryear{Amiri et~al.,}{Amiri et~al.}{2019a}]{chime400}
Amiri M.,  et~al., 2019a, \mn@doi [Nature] {10.1038/s41586-018-0867-7}, 566,
  230

\bibitem[\protect\citeauthoryear{Amiri et~al.,}{Amiri et~al.}{2019b}]{R2}
Amiri M.,  et~al., 2019b, \mn@doi [Nature] {10.1038/s41586-018-0864-x}, 566,
  235

\bibitem[\protect\citeauthoryear{{Bannister} et~al.,}{{Bannister}
  et~al.}{2017}]{ASKAP}
{Bannister} K.~W.,  et~al., 2017, \mn@doi [\apjl] {10.3847/2041-8213/aa71ff},
  \href {http://adsabs.harvard.edu/abs/2017ApJ...841L..12B} {841, L12}

\bibitem[\protect\citeauthoryear{{Bhandari} et~al.,}{{Bhandari}
  et~al.}{2018}]{Bhandari2018}
{Bhandari} S.,  et~al., 2018, \mn@doi [\mnras] {10.1093/mnras/stx3074}, \href
  {http://adsabs.harvard.edu/abs/2018MNRAS.475.1427B} {475, 1427}

\bibitem[\protect\citeauthoryear{{Bhattacharyya} et~al.,}{{Bhattacharyya}
  et~al.}{2016}]{Bhattacharyya2016}
{Bhattacharyya} B.,  et~al., 2016, \mn@doi [\apj]
  {10.3847/0004-637X/817/2/130}, \href
  {http://adsabs.harvard.edu/abs/2016ApJ...817..130B} {817, 130}

\bibitem[\protect\citeauthoryear{{Bilicki} et~al.,}{{Bilicki}
  et~al.}{2016}]{Bilicki2016}
{Bilicki} M.,  et~al., 2016, \mn@doi [\apjs] {10.3847/0067-0049/225/1/5}, \href
  {http://adsabs.harvard.edu/abs/2016ApJS..225....5B} {225, 5}

\bibitem[\protect\citeauthoryear{{Chatterjee} et~al.,}{{Chatterjee}
  et~al.}{2017}]{VLAlocalisation}
{Chatterjee} S.,  et~al., 2017, \mn@doi [\nat] {10.1038/nature20797}, \href
  {http://adsabs.harvard.edu/abs/2017Natur.541...58C} {541, 58}

\bibitem[\protect\citeauthoryear{{Connor}, {Pen}  \& {Oppermann}}{{Connor}
  et~al.}{2016}]{2016MNRAS.458L..89C}
{Connor} L.,  {Pen} U.-L.,   {Oppermann} N.,  2016, \mn@doi [\mnras]
  {10.1093/mnrasl/slw026}, \href
  {http://adsabs.harvard.edu/abs/2016MNRAS.458L..89C} {458, L89}

\bibitem[\protect\citeauthoryear{{Cordes} \& {Lazio}}{{Cordes} \&
  {Lazio}}{2002}]{Cordes}
{Cordes} J.~M.,  {Lazio} T.~J.~W.,  2002, ArXiv Astrophysics e-prints
  arXiv:astro-ph/0207156, \href
  {http://adsabs.harvard.edu/abs/2002astro.ph..7156C} {}

\bibitem[\protect\citeauthoryear{{Deneva} et~al.,}{{Deneva}
  et~al.}{2009}]{Deneva2009}
{Deneva} J.~S.,  et~al., 2009, \mn@doi [\apj] {10.1088/0004-637X/703/2/2259},
  \href {http://adsabs.harvard.edu/abs/2009ApJ...703.2259D} {703, 2259}

\bibitem[\protect\citeauthoryear{{Farah} et~al.,}{{Farah}
  et~al.}{2018}]{Farah2018}
{Farah} W.,  et~al., 2018, \mn@doi [\mnras] {10.1093/mnras/sty1122}, \href
  {http://adsabs.harvard.edu/abs/2018MNRAS.478.1209F} {478, 1209}

\bibitem[\protect\citeauthoryear{{Gajjar} et~al.,}{{Gajjar}
  et~al.}{2018}]{2018ApJ...863....2G}
{Gajjar} V.,  et~al., 2018, \mn@doi [\apj] {10.3847/1538-4357/aad005}, \href
  {http://adsabs.harvard.edu/abs/2018ApJ...863....2G} {863, 2}

\bibitem[\protect\citeauthoryear{{Hay} \& {O'Sullivan}}{{Hay} \&
  {O'Sullivan}}{2008}]{Hay2008}
{Hay} S.~G.,  {O'Sullivan} J.~D.,  2008, \mn@doi [Radio Science]
  {10.1029/2007RS003798}, \href
  {http://adsabs.harvard.edu/abs/2008RaSc...43.6S04H} {43, RS6S04}

\bibitem[\protect\citeauthoryear{{Hotan} et~al.,}{{Hotan}
  et~al.}{2014}]{Hotan2014}
{Hotan} A.~W.,  et~al., 2014, \mn@doi [PASA] {10.1017/pasa.2014.36}, \href
  {http://adsabs.harvard.edu/abs/2014PASA...31...41H} {31, e041}

\bibitem[\protect\citeauthoryear{{Ioka}}{{Ioka}}{2003}]{Ioka}
{Ioka} K.,  2003, \mn@doi [ApJL] {10.1086/380598}, \href
  {http://adsabs.harvard.edu/abs/2003ApJ...598L..79I} {598, L79}

\bibitem[\protect\citeauthoryear{{James}}{{James}}{2019}]{James2019}
{James} C.~W.,  2019, preprint, \href
  {http://adsabs.harvard.edu/abs/2019arXiv190204932J} {} (\mn@eprint {arXiv}
  {1902.04932})

\bibitem[\protect\citeauthoryear{{James} et~al.,}{{James}
  et~al.}{2018}]{2018arXiv181004356J}
{James} C.~W.,  et~al., 2018, preprint, \href
  {http://adsabs.harvard.edu/abs/2018arXiv181004356J} {} (\mn@eprint {arXiv}
  {1810.04356})

\bibitem[\protect\citeauthoryear{{Keane} et~al.,}{{Keane}
  et~al.}{2018}]{SUPERB1}
{Keane} E.~F.,  et~al., 2018, \mn@doi [\mnras] {10.1093/mnras/stx2126}, \href
  {http://adsabs.harvard.edu/abs/2018MNRAS.473..116K} {473, 116}

\bibitem[\protect\citeauthoryear{{Keith} et~al.,}{{Keith} et~al.}{2010}]{BPSR}
{Keith} M.~J.,  et~al., 2010, \mn@doi [\mnras]
  {10.1111/j.1365-2966.2010.17325.x}, \href
  {http://adsabs.harvard.edu/abs/2010MNRAS.409..619K} {409, 619}

\bibitem[\protect\citeauthoryear{{Law} et~al.,}{{Law}
  et~al.}{2017}]{2017ApJ...850...76L}
{Law} C.~J.,  et~al., 2017, \mn@doi [\apj] {10.3847/1538-4357/aa9700}, \href
  {http://adsabs.harvard.edu/abs/2017ApJ...850...76L} {850, 76}

\bibitem[\protect\citeauthoryear{{Macquart} et~al.,}{{Macquart}
  et~al.}{2010}]{mbb+10}
{Macquart} J.-P.,  et~al., 2010, PASA, 27, 272

\bibitem[\protect\citeauthoryear{{Mahony} et~al.,}{{Mahony}
  et~al.}{2018}]{Mahony2018}
{Mahony} E.~K.,  et~al., 2018, \mn@doi [\apjl] {10.3847/2041-8213/aae7cb},
  \href {http://adsabs.harvard.edu/abs/2018ApJ...867L..10M} {867, L10}

\bibitem[\protect\citeauthoryear{{Manchester} et~al.,}{{Manchester}
  et~al.}{2001}]{Manchester2001}
{Manchester} R.~N.,  et~al., 2001, \mn@doi [\mnras]
  {10.1046/j.1365-8711.2001.04751.x}, \href
  {http://adsabs.harvard.edu/abs/2001MNRAS.328...17M} {328, 17}

\bibitem[\protect\citeauthoryear{{McQuinn}}{{McQuinn}}{2014}]{Quinn}
{McQuinn} M.,  2014, \mn@doi [ApJL] {10.1088/2041-8205/780/2/L33}, \href
  {http://adsabs.harvard.edu/abs/2014ApJ...780L..33M} {780, L33}

\bibitem[\protect\citeauthoryear{{Michilli} et~al.,}{{Michilli}
  et~al.}{2018}]{Michilli2018}
{Michilli} D.,  et~al., 2018, \mn@doi [\nat] {10.1038/nature25149}, \href
  {http://adsabs.harvard.edu/abs/2018Natur.553..182M} {553, 182}

\bibitem[\protect\citeauthoryear{{Oppermann}, {Yu}  \& {Pen}}{{Oppermann}
  et~al.}{2018}]{Oppermann2018}
{Oppermann} N.,  {Yu} H.-R.,   {Pen} U.-L.,  2018, \mn@doi [\mnras]
  {10.1093/mnras/sty004}, \href
  {http://adsabs.harvard.edu/abs/2018MNRAS.475.5109O} {475, 5109}

\bibitem[\protect\citeauthoryear{{Patel} et~al.,}{{Patel}
  et~al.}{2018}]{Patel2018}
{Patel} C.,  et~al., 2018, \mn@doi [\apj] {10.3847/1538-4357/aaee65}, \href
  {http://adsabs.harvard.edu/abs/2018ApJ...869..181P} {869, 181}

\bibitem[\protect\citeauthoryear{{Petroff} et~al.,}{{Petroff}
  et~al.}{2016}]{pbj+16}
{Petroff} E.,  et~al., 2016, PASA, 33, e045

\bibitem[\protect\citeauthoryear{{Planck Collaboration} et~al.,}{{Planck
  Collaboration} et~al.}{2016}]{2016A&A...594A..13P}
{Planck Collaboration} et~al., 2016, \mn@doi [\aap]
  {10.1051/0004-6361/201525830}, \href
  {http://adsabs.harvard.edu/abs/2016A%26A...594A..13P} {594, A13}

\bibitem[\protect\citeauthoryear{{Price} et~al.,}{{Price}
  et~al.}{2018}]{2018RNAAS...2a..30P}
{Price} D.~C.,  et~al., 2018, \mn@doi [Research Notes of the American
  Astronomical Society] {10.3847/2515-5172/aaaf69}, \href
  {http://adsabs.harvard.edu/abs/2018RNAAS...2a..30P} {2, 30}

\bibitem[\protect\citeauthoryear{{Ravi} et~al.,}{{Ravi}
  et~al.}{2016}]{RaviScience}
{Ravi} V.,  et~al., 2016, \mn@doi [Science] {10.1126/science.aaf6807}, \href
  {http://adsabs.harvard.edu/abs/2016Sci...354.1249R} {354, 1249}

\bibitem[\protect\citeauthoryear{{Scholz} et~al.,}{{Scholz}
  et~al.}{2016}]{ssh+16b}
{Scholz} P.,  et~al., 2016, \mn@doi [\apj] {10.3847/1538-4357/833/2/177}, \href
  {http://adsabs.harvard.edu/abs/2016ApJ...833..177S} {833, 177}

\bibitem[\protect\citeauthoryear{{Shannon} et~al.,}{{Shannon}
  et~al.}{2018}]{Shannon2018}
{Shannon} R.~M.,  et~al., 2018, \mn@doi [\nat] {10.1038/s41586-018-0588-y},
  \href {http://adsabs.harvard.edu/abs/2018Natur.562..386S} {562, 386}

\bibitem[\protect\citeauthoryear{{Spitler} et~al.,}{{Spitler}
  et~al.}{2014}]{sch+14}
{Spitler} L.~G.,  et~al., 2014, ApJ, 790, 101

\bibitem[\protect\citeauthoryear{{Spitler} et~al.,}{{Spitler}
  et~al.}{2016}]{ssh+16a}
{Spitler} L.~G.,  et~al., 2016, Nature, 531, 202

\bibitem[\protect\citeauthoryear{{Spitler} et~al.,}{{Spitler}
  et~al.}{2018}]{2018ApJ...863..150S}
{Spitler} L.~G.,  et~al., 2018, \mn@doi [\apj] {10.3847/1538-4357/aad332},
  \href {http://adsabs.harvard.edu/abs/2018ApJ...863..150S} {863, 150}

\bibitem[\protect\citeauthoryear{{Stovall} et~al.,}{{Stovall}
  et~al.}{2014}]{Stovall2014}
{Stovall} K.,  et~al., 2014, \mn@doi [\apj] {10.1088/0004-637X/791/1/67}, \href
  {http://adsabs.harvard.edu/abs/2014ApJ...791...67S} {791, 67}

\bibitem[\protect\citeauthoryear{{Tendulkar} et~al.,}{{Tendulkar}
  et~al.}{2017}]{Host}
{Tendulkar} S.~P.,  et~al., 2017, \mn@doi [\apjl] {10.3847/2041-8213/834/2/L7},
  \href {http://adsabs.harvard.edu/abs/2017ApJ...834L...7T} {834, L7}

\bibitem[\protect\citeauthoryear{{Walker}, {Ma}  \& {Breton}}{{Walker}
  et~al.}{2018}]{Walker2018}
{Walker} C.~R.~H.,  {Ma} Y.-Z.,   {Breton} R.~P.,  2018, preprint, \href
  {http://adsabs.harvard.edu/abs/2018arXiv180401548W} {} (\mn@eprint {arXiv}
  {1804.01548})

\bibitem[\protect\citeauthoryear{{Wright} et~al.,}{{Wright}
  et~al.}{2010}]{Wright2010}
{Wright} E.~L.,  et~al., 2010, \mn@doi [\aj] {10.1088/0004-6256/140/6/1868},
  \href {http://adsabs.harvard.edu/abs/2010AJ....140.1868W} {140, 1868}

\bibitem[\protect\citeauthoryear{{Xu} \& {Han}}{{Xu} \& {Han}}{2015}]{Xu2015}
{Xu} J.,  {Han} J.~L.,  2015, \mn@doi [Research in Astronomy and Astrophysics]
  {10.1088/1674-4527/15/10/002}, \href
  {http://adsabs.harvard.edu/abs/2015RAA....15.1629X} {15, 1629}

\bibitem[\protect\citeauthoryear{{Yao}, {Manchester}  \& {Wang}}{{Yao}
  et~al.}{2017}]{YMW16}
{Yao} J.~M.,  {Manchester} R.~N.,   {Wang} N.,  2017, \mn@doi [\apj]
  {10.3847/1538-4357/835/1/29}, \href
  {http://adsabs.harvard.edu/abs/2017ApJ...835...29Y} {835, 29}

\bibitem[\protect\citeauthoryear{{Zackay} \& {Ofek}}{{Zackay} \&
  {Ofek}}{2017}]{Zackay2017}
{Zackay} B.,  {Ofek} E.~O.,  2017, \mn@doi [\apj] {10.3847/1538-4357/835/1/11},
  \href {http://adsabs.harvard.edu/abs/2017ApJ...835...11Z} {835, 11}

\bibitem[\protect\citeauthoryear{{Zhang}, {Gajjar}, {Foster}, {Siemion},
  {Cordes}, {Law}  \& {Wang}}{{Zhang} et~al.}{2018}]{2018ApJ...866..149Z}
{Zhang} Y.~G.,  {Gajjar} V.,  {Foster} G.,  {Siemion} A.,  {Cordes} J.,  {Law}
  C.,   {Wang} Y.,  2018, \mn@doi [\apj] {10.3847/1538-4357/aadf31}, \href
  {http://adsabs.harvard.edu/abs/2018ApJ...866..149Z} {866, 149}

\bibitem[\protect\citeauthoryear{{Zhou}, {Li}, {Wang}, {Fan}  \& {Wei}}{{Zhou}
  et~al.}{2014}]{Zhou2014}
{Zhou} B.,  {Li} X.,  {Wang} T.,  {Fan} Y.-Z.,   {Wei} D.-M.,  2014, \mn@doi
  [\prd] {10.1103/PhysRevD.89.107303}, \href
  {http://adsabs.harvard.edu/abs/2014PhRvD..89j7303Z} {89, 107303}

\makeatother
\end{thebibliography}



\bsp	
\label{lastpage}
\end{document}